\documentclass[11pt]{article}
\usepackage{amssymb,amsfonts,amsmath}
\usepackage{enumerate,float,comment}
\usepackage{color}
\usepackage{tikz}
\usetikzlibrary{backgrounds}
\usepackage{xcolor}
\definecolor{mintcream}{rgb}{0.96, 1.0, 0.98}
\definecolor{champagne}{rgb}{0.97, 0.91, 0.81}
\definecolor{bubblegum}{rgb}{0.99, 0.76, 0.8}
\definecolor{airforceblue}{rgb}{0.36, 0.54, 0.66}
\definecolor{persianblue}{rgb}{0.11, 0.22, 0.73}
\definecolor{zaffre}{rgb}{0.0, 0.08, 0.66}
\usepackage[pdftex,unicode,
colorlinks=true,
linkcolor = zaffre,
citecolor = cyan,
urlcolor = airforceblue]{hyperref}
\usepackage{cleveref}
\usepackage{tcolorbox}

\usetikzlibrary{arrows,snakes,backgrounds}

\def\be{\begin{eqnarray}}
\def\ee{\end{eqnarray}}

\definecolor{red}{rgb}{1,0,0}
\definecolor{orange}{rgb}{1,0.5,0}
\definecolor{violet}{rgb}{0.7,0,1}

\usepackage[top = 2 cm, bottom = 2 cm, left = 2.5 cm, right = 1.5 cm]{geometry}

\usepackage[utf8]{inputenc}

\usepackage[natbib=true, backend = bibtex, style=numeric-comp, sorting=none]{biblatex}
\begin{document}

\hfill MIPT/TH-23/26

\hfill IITP/TH-25/26

\hfill ITEP/TH-28/26

\bigskip

\centerline{\Large{
Towards  Equations  for  String  Amplitudes
}}

\bigskip
 
\bigskip
 
\centerline{\bf A.Morozov$^{a,b,c,d,}$\footnote{morozov@itep.ru}, K.Pushkin$^{a,d,}$\footnote{pushkin.ks@phystech.edu} and M.Reva$^{a,b,d,}$\footnote{reva.ma@phystech.edu}}
 
\bigskip
 
\centerline{$^a$ {\it MIPT, Dolgoprudny, 141701, Russia}}
\centerline{$^b$ {\it NRC ``Kurchatov Institute", 123182, Moscow, Russia}}
\centerline{$^c$ {\it Institute for Information Transmission Problems, Moscow 127994, Russia}}
\centerline{$^d$ {\it ITEP, 117259, Moscow, Russia}}
 
\bigskip
 
\bigskip
 
\centerline{ABSTRACT}

\bigskip
 
{\footnotesize
 Generic Feynman integrals are widely studied as solutions of Picard-Fuchs equations 
on moduli spaces of their parameters, and this calls for consideration of this phenomenon
at a more basic level -- of string amplitudes which are integrals over true non-singular
module space of Riemann surfaces and their various generalizations.
The main puzzle here is that a single string amplitude involves mane different
particle diagrams, corresponding to different parts of the same moduli space,
but different particle diagrams are usually believed to satisfy different equations,
not unified into a common entity.
We begin investigation of this problem, starting from  Koba-Nielsen diagrams.
While there is nothing interesting at this level for particles,
the tree-level open bosonic string amplitudes satisfy non-trivial linear difference equations in kinematic variables.
Moreover, the  integration‑by‑parts  on moduli space, standing behind Picard-Fuchs equations for particle loops,
for strings are operative already at the tree level.
We construct a complete system of such equations for arbitrary
n-point tree amplitudes, 
with the number of independent relations matching the kinematic parameters. 
In variance with the particle case equations are difference ones rather than differential.
The low‑energy limit $\alpha\to 0$ smoothly recovers the algebraic QFT structure.
}

\bigskip

\bigskip

\bigskip

\section{Introduction}

Scattering amplitudes in quantum field theory (QFT) can be expanded into sums of Feynman diagrams,
and these diagrams are known to satisfy certain equations w.r.t. external momenta \cite{weinzierl2022feynmanintegrals}.
These equations depend on the diagram and get especially interesting in the multiloop case,
when they are actually the Picard-Fucks equations for (Feynman) integrals, depending on parameters.
Specific for Feynman integrals are simplifications after Fourier transform from
momentum to configuration space \cite{Mishnyakov:2024rmb,Mishnyakov:2023sly,Mishnyakov:2024xjz,Diakonov:2024qjn}.    
The problem is still far from complete understanding and even classification, 
but it is under intensive study with a lot of impressive results.

In this paper we raise the question of generalization from particles to strings --
which still attracted much less attention.
The Feynman calculus for strings is well developed \cite{Polyakov:1987hqn,Morozov:1992ea,RevModPhys.60.917,Belavin:1986tv,Knizhnik:1989ak,Morozov:1989cw}, 
and integrals are now over moduli spaces of Riemann surfaces,
with peculiar holomorphic measures, which are bilinear combinations of holomorphic ones.
In the simplest case of bosonic string in 26 space-time dimensions this is
just the square modulus of the celebrated Mumford measure \cite{Belavin:1986cy,Beilinson:1986zw}, 
while in general measures involve correlators in various 2d conformal field theories (CFT).    
Thus classification of equations, if any, would include the interplay of varieties
of Riemann surfaces and CFT's and should be far more interesting and better structured 
than in the case of particles,
where these entities are not separated and mixed in the tropical limit,
substituting surfaces by diagrams.

In this introductory note we do not go deep into this story and instead focus on the 
simplest string amplitudes with the goal to emphasize the initial difference 
from the particle case.
Perhaps, the main one is that a single string amplitude is a linear combination
of many particle ones.
This has two essentially different reasons.
One is that string is a collection of infinitely many (tower of) particles of different
masses and spins.
Another is that a Riemann surface interpolates between several different tropical limits,
i.e. the moduli space has several different particle diagrams at its boundary.
Say the 4-point tree amplitudes in the $s$, $t$ and $u$ channels belong to one and the
same string amplitude (Veneziano amplitude).
At the same time the peculiar feature of {\it duality} forces consideration 
of just one channel instead of the sum over three -- and the net result 
after summation over the tower of states is independent of the choice of the channel.
These features makes the string/particle interplay somewhat obscure. 
Even if one has already some intuition about equations in particle case,
it is not straightforwardly extended to string level.
And the backward problem of reduction from strings to particles is also not fully
straightforward and not exhausted by the reference to $\alpha'\longrightarrow 0$ limit \cite{Schwarz:1973yz,SCHERK1971222,Dearnaley:1989dc} --
at the level of equations the limit needs care and more detailed description. 

In this paper we concentrate on just the derivation and description of equations {\it per se}
in the simplest examples -- of tree sting amplitudes.
All above-mentioned subjects should be discussed at the next stages,
after the equations themselves get established and verified by the community.
  
\section{Veneziano amplitude}
In this chapter, we consider the scattering process of an open string. One of the main objects of consideration is the four-point Veneziano amplitude \cite{Mandelstam:1968czc}.
\begin{equation}
    A_4(\alpha s,\alpha t)=\int_0^1dx x^{-\alpha s-2}(1-x)^{-\alpha t-2}=\frac{\Gamma(-\alpha s-1)\Gamma(-\alpha t-1)}{\Gamma(-\alpha s-\alpha t-2)}
\end{equation}
where $s=-(p_1+p_2)^2,t=-(p_1+p_3)^2$.\\
One of the properties of the Veneziano amplitude is the fact that it is expressed in terms of the Euler gamma functions that satisfy difference equations of the form:
\begin{equation}
\begin{aligned}
    (D_z-&(z-1))\Gamma(z)=0\\
    D_z f(z)& = f(z+1)-f(z)
\end{aligned}
\end{equation}
Later on we will call $D_z$ the difference operator.\\
Using this property of gamma functions, it is possible to construct a difference equation for the Veneziano amplitude
\begin{equation}
    ((2+\alpha s)D_{\alpha s}-(1+\alpha t))A_4(\alpha s,\alpha t)=0
\end{equation}
\begin{equation}
    ((2+\alpha t)D_{\alpha t}-(1+\alpha s))A_4(\alpha s,\alpha t)=0
\end{equation}
The resulting equations have the properties of duality ($s \leftrightarrow t$), which is not astonishing, since the amplitude of string theory has such a property. One of the interesting observations of the Veneziano amplitude is the fact that it is impossible to write a homogeneous linear differential equation in one variable with rational coefficients on it, as it has infinite number of poles and its derivative's divergence degree will grow, compared to initial one, in the vicinity of poles, not allowing to find such polynomial coefficients, that will cancel out the growth of pole order. The simplest example of this phenomena can be obtained from analyzing $\tan{x}$. This function has poles at the points $x = \frac{\pi}{2} + \mathbb{Z}\pi i$ of first order. However, its derivatives have poles at the exactly same points, but of increasingly higher order:
\begin{equation}
    \frac{\partial}{\partial x} \tan{x} = 1 + \tan^2{x} \rightarrow \frac{\partial^n}{\partial x^n} \tan{x} \sim \backslash \text{neighborhood of poles} \backslash \sim \tan^{n+1}{x}
\end{equation}
Thus we need to find polynomial coefficients for ODE which will have infinite number of zeros to cancel out rising pole order. However, any one variable polynomial have finite number of roots, thus forbidding the construction of such equation. Actually, this phenomena acquires in amplitudes of higher order, preventing us from naive attempt to generalize well known equations on QFT diagrams \cite{Mishnyakov:2024rmb,Mishnyakov:2023sly,Mishnyakov:2024xjz,Lairez:2022zkj,Bonisch:2020qmm,Duhr:2025ouy,Levkovich-Maslyuk:2024zdy} to the string case.\\
In order to verify our equations, we will use the fact that in the limit of high energies (Regge's limit), i.e. $s\to \infty$ and fixed t, the amplitude has a power dependence on $s$ \cite{Veneziano:1968yb}:\
\begin{equation}
    A(\alpha s,\alpha t) \sim (\alpha s)^{1+\alpha t}
\end{equation}
In this limit our equations turn into differential ones, since the difference operator becomes the differentiation operator. The obtained equations correspond and are identical for the amplitude behavior according to the power law.
\begin{equation}
\begin{aligned}
    (\alpha s\partial_{\alpha s}-(1+\alpha t))A_4^{s\to \infty}=0\\
    (\alpha t\partial_{\alpha t}-(1+\alpha s))A_4^{t\to\infty}=0
\end{aligned}
\end{equation}
After making sure that our equations are correct, we can try to find a solution to the difference equations. \\
In fact, we can decompose the Veneziano amplitude into an infinite number of tree diagrams along the s-channel.
\begin{equation}
    A(\alpha s,\alpha t)=-\sum_{n=0}^{\infty}\frac{(\alpha t+2)(\alpha t+3)\dots(\alpha t+1+n)}{n!}\frac{1}{\alpha s+1-n}
\end{equation}
Knowing that the amplitude can be decomposed into an s-channel, we can try to find a solution to the difference equation using a formal series $A(\alpha s,\alpha t)=-\sum_n\frac{a_n(\alpha t)}{\alpha s+1-n}$.\\
Substituting this series into our equations, we obtain a recurrence relation for the coefficients $a_n(\alpha t)$:
\begin{equation}
\begin{aligned}
    (2+\alpha s)&\Big(\sum_{n=0}^{\infty}\frac{a_n}{\alpha s+2-n}-\sum_{n=0}^{\infty}\frac{a_n}{\alpha s+1-n}\Big)-(1+\alpha t)\sum_{n=0}^{\infty}\frac{a_n}{\alpha s+1-n}=\\
    & = \sum_{n=0}^{\infty}a_n+\sum_{n=0}^{\infty}\frac{a_n n}{\alpha s+2-n} -\sum_{n=0}^{\infty}a_n-\sum_{n=0}^{\infty}\frac{a_n(1+n)}{\alpha s+1-n}-(1+\alpha t)\sum_{n=0}^{\infty}\frac{a_n}{\alpha s+1-n}=0 
\end{aligned}
\end{equation}
By making a shift in the second and third series $n\to n-1$ and use ansatz $a_0=1, a_i=0 (i<0)$ we get:
\begin{equation}
    \sum_{n=0}^{\infty}\frac{na_n}{\alpha s+2-n}-\sum_{n=0}^{\infty}\frac{n a_{n-1}}{\alpha s+2-n}-(1+\alpha t)\sum_{n=0}^{\infty}\frac{a_{n-1}}{\alpha s+2-n}=0
\end{equation}
Here we used the fact that:
\begin{equation}
    \sum_{n=0}^{\infty}\frac{a_n(n+1)}{\alpha s+1-n}\to (n\to n-1)\to \sum_{n=0}^{\infty}\frac{a_{n-1}n}{\alpha s+2-n}
\end{equation}
As a result, we get a recurrent equation that we can solve:
\begin{equation}
    a_nn-a_{n-1}n-(1+\alpha t)a_{n-1}=0
\end{equation}
\begin{equation}
    a_{n+1}=\frac{(2+\alpha t+n)}{n+1}a_n
\end{equation}
As a result we get a solution based on coefficients, which coincides with the coefficient in formula (8). Summing up all the calculations in this chapter, we can draw several unique conclusions, namely, we understand that the string amplitude satisfies difference equations, which distinguishes it from the amplitude in QFT, where the tree diagram in momentum space satisfies differential equations, where we will discuss this in detail in the next chapter. The resulting equations also describe the behavior of the amplitude well at high energy, but the most truly amazing property of the equations is the fact that it can be solved as a formal series using $s$-channel expansion.

\section{Tree-level QFT}
In this section, we discuss the equations in coordinate and momentum space for tree-level amplitudes in the QFT. In the coordinate space, the amplitude of the tree-diagram is written in a simple form as an integral containing Green's functions that connect the vertices of the diagram, while integration occurs only along the inner lines. The Green's scalar function satisfies the Klein-Gordon differential equations, and in momentum space, after Fourier transformation, this function satisfies an algebraic equation, namely
\begin{equation}
\begin{aligned}
    &\text{coordinate:}\hspace{7.5mm} (\Box+m^2) G(x,y) = \delta(x-y)\\
    &\text{momentum:}\hspace{7mm} (p^2-m^2) \tilde G(P) = 1
\end{aligned}
\end{equation}
Unlike in QFT, where individual Feynman diagrams are typically used, in string theory, a single string diagram already contains the contribution of an infinite number of such diagrams corresponding to different intermediate states $m^2_n$. Therefore, it is natural to move on to discussing equations not for individual diagrams, but directly for the complete scattering.\\
Before moving on to the more general case of amplitudes containing multiple channels, it is useful to discuss the simplest configurations in detail. In particular, we will focus on cases where the amplitude contains only a single massless channel, which allows us to illustrate the main features of the amplitude structure and prepare the ground for further generalization.
\subsection{Four-point amplitude} Let's start by considering the s-channel tree diagram in coordinate space. The corresponding amplitude can be written as:
\begin{equation}
   A(x_1,\dots,x_4)=G(x_3-x_1)\delta^{4}(x_2-x_1)\delta^{4}(x_3-x_4)
\end{equation}
Where $G(x-y)$ is the Green's function of a scalar field. This structure reflects the locality of interactions at the vertices of the diagram and the propagation of the intermediate state.\\
Using the properties of the Green's function, it is easy to show that this amplitude satisfies the differential equation
\begin{equation}
(\partial_{x_1}+\partial_{x_2})^2A(x_1,\dots,x_4)=\delta^4(x_3-x_1)\delta^4(x_2-x_1)\delta^4(x_3-x_4)
\end{equation}
This equation expresses the fact that the kinetic energy operator acts on the intermediate line of the diagram, resulting in a contact source.\\
We now pass to the impulse representation by performing a Fourier transform. In this case, the differential operator is replaced by an algebraic multiplier, and the amplitude satisfies the equation
\begin{equation}
    (p_1+p_2)^2A(p_1,\dots,p_4)=\delta^4(p_4+p_3-p_2-p_1)\to A(p_1,\dots,p_4)=\frac{\delta^4(p_4+p_3-p_2-p_1)}{(p_1+p_2)^2}
\end{equation}
This ratio demonstrates a characteristic feature of tree diagrams: the amplitude is determined by an algebraic operator associated with an intermediate propagator.\\
A more physically meaningful case arises when all three scattering channels — s, t, and u — are taken into account. In this case, the total tree amplitude is given by the sum of the corresponding contributions:
\begin{equation}
\begin{aligned}
    A(x_1,\dots,x_4)=G(x_3-x_1)\delta^4(x_2&-x_1)\delta^4(x_3-x_4) + G(x_2-x_1)\delta^4(x_2-x_3)\delta^4(x_1-x_4) + \\ + &G(x_2-x_1)\delta^4(x_1-x_3)\delta^4(x_2-x_4)
\end{aligned}
\end{equation}
The sum over three-channels satisfies a differential equation in coordinate space
\begin{equation}
\begin{aligned}
    (\partial_{x_2}+\partial_{x_2})^2&(\partial_{x_1}-\partial_{x_3})^2(\partial_{x_1}-\partial_{x_4})^2A(x_1,\dots,x_4) = \\ & (\partial_{x_1}-\partial_{x_3})^2(\partial_{x_1}-\partial_{x_4})^2(\delta^4(x_1-x_2)\delta^4(x_2-x_3)\delta^4(x_2-x_4))+ \\
    + & (\partial_{x_1}+\partial_{x_2})^2(\partial_{x_1}-\partial_{x_4})^2(\delta^4(x_1-x_3)\delta^4(x_2-x_3)\delta^4(x_3-x_4)) + \\ 
    + & (\partial_{x_1}+\partial_{x_2})^2(\partial_{x_1}-\partial_{x_3})^2(\delta^4(x_1-x_4)\delta^4(x_2-x_4)\delta^4(x_4-x_3)
\end{aligned}
\end{equation}
As a result, the total amplitude obeys a system of differential equations in coordinate space, which, after Fourier transformation, becomes an algebraic relation between the impulses.
\begin{equation}
\begin{aligned}
    (p_1+p_2)^2(p_1-p_3)^2(p_1-&p_4)^2A(p_1,\dots,p_4)=\delta^4(p_4+p_3-p_2-p_1)\big((p_1-p_3)^2(p_1-p_4)^2+ \\
    &+(p_1+p_2)^2(p_1-p_4)^2+(p_1+p_2)^2(p_1-p_3)^2\big)
\end{aligned}
\end{equation}
Thus, even in the simplest four-point case, it is seen that the tree amplitudes of quantum field theory are naturally described by equations of motion — differential in coordinate space and algebraic in momentum space. This structure serves as a starting point for further generalization to more complex processes and for comparison with the equations that arise in string theory.
\subsection{Five-point amplitude}
After analyzing the four-point tree amplitude, the natural next step is to consider the five-point process. This case already demonstrates a significantly richer kinematic structure and allows us to observe how the equations for amplitudes are generalized as the number of external states increases.\\
The amplitude in the coordinate space reads:
\begin{equation}
    \begin{aligned}
        A_5(x_1,...,x_5)=&G(x_1-x_5)G(x_2-x_5)\delta^4(x_2-x_3)\delta^4(x_4-x_2) + \\
        +&G(x_1-x_2)G(x_1-x_3)\delta^4(x_4-x_3)\delta^4(x_2-x_5)+ \\
        +&G(x_2-x_1)G(x_2-x_3)\delta^4(x_4-x_3)\delta^4(x_1-x_5)+ \\
        +&G(x_1-x_4)G(x_4-x_3)\delta^4(x_1-x_3)\delta^4(x_3-x_5)+ \\
        +&G(x_1-x_3)G(x_3-x_4)\delta^4(x_1-x_2)\delta^4(x_4-x_5)
    \end{aligned}
\end{equation}
where we have natural decomposition into independent channels.\\
Thus it satisfies the following differential equation:
\begin{equation}
\begin{aligned}
  &(\partial_{x_1}-\partial_{x_2})^2(\partial_{x_4}-\partial_{x_3})^2(\partial_{x_2}+\partial_{x_5})^2(\partial_{x_1}-\partial_{x_5})^2(\partial_{x_3}+\partial_{x_5})^2(\partial_{x_4}-\partial_{x_5})^2A(x_1,..,x_5)= \\
  &\hspace{5mm} (\partial_{x_1}+\partial_{x_4}-\partial_{x_3})^2(\partial_{x_3}+\partial_{x_2}-\partial_{x_4})^2(\partial_{x_1}+\partial_{x_4}-\partial_{x_2})^2(\partial_{x_3}+\partial_{x_2}-\partial_{x_1})\times \\
  & \hspace{70mm} \times \delta^4(x_1-x_5)\delta^4(x_4-x_5)\delta^4(x_5-x_2)\delta^4(x_5-x_3) + \\
  & \hspace{5mm} + (\partial_{x_1}-\partial_{x_2})^2(\partial_{x_3}+\partial_{x_2}-\partial_{x_4})^2(\partial_{x_1}+\partial_{x_4}-\partial_{x_2})^2(\partial_{x_2}+\partial_{x_3}-\partial_{x_1})^2 \times \\
  & \hspace{70mm}\times  \delta^4(x_1-x_5)\delta^4(x_4-x_5)\delta^4(x_5-x_2)\delta^{4}(x_5-x_3) + \\
  & \hspace{5mm} + (\partial_{x_1}-\partial_{x_2})^2(\partial_{x_1}+\partial_{x_4}-\partial_{x_3})^2(\partial_{x_1}+\partial_{x_4}-\partial_{x_2})^2(\partial_{x_3}+\partial_{x_2}-\partial_{x_1})^2 \times \\
  & \hspace{70mm} \delta^4(x_1-x_5)\delta^4(x_4-x_5)\delta^4(x_5-x_2)\delta^4(x_5-x_3) + \\
  & \hspace{5mm} + (\partial_{x_4}-\partial_{x_3})^2(\partial_{x_1}+\partial_{x_4}-\partial_{x_3})^2(\partial_{x_3}+\partial_{x_2}-\partial_{x_4})^2(\partial_{x_3}+\partial_{x_2}-\partial_{x_1})^2 \times \\
  & \hspace{70mm} \times \delta^4(x_1-x_5)\delta^4(x_4-x_5)\delta^4(x_5-x_2)\delta^4(x_5-x_3) + \\
  & \hspace{5mm} + (\partial_{x_4}-\partial_{x_3})^2(\partial_{x_1}+\partial_{x_4}-\partial_{x_3})^2(\partial_{x_3}+\partial_{x_2}-\partial_{x_4})^2(\partial_{x_1}+\partial_{x_4}-\partial_{x_2})^2 \times \\
  & \hspace{70mm} \times (\delta^4(x_1-x_5)\delta^4(x_4-x_5)\delta^4(x_5-x_2)\delta^4(x_5-x_3)  \\
\end{aligned}
\end{equation}
Moving further to the momentum space we get the desired algebraic equation
\begin{equation}
    \begin{aligned}
 &(p_1-p_2)^2(p_4-p_3)^2(p_2+p_5)^2(p_1-p_5)^2(p_3+p_5)^2(p_4-p_5)^2A(p_1,..,p_5) = \\ 
 &\delta^4(p_1+p_4-p_3-p_2-p_5) \times\\
 &\times \Big((p_2+p_5)^2(p_1-p_5)^2(p_3+p_5)^2(p_4-p_5)^2+(p_1-p_2)^2(p_1-p_5)^2(p_3+p_5)^2(p_4-p_5)^2+\\
 &\phantom{\times \Big( (} (p_1-p_2)^2(p_2+p_5)^2(p_3+p_5)^2(p_4-p_5)^2+(p_4-p_3)^2(p_2+p_5)^2(p_1-p_5)^2(p_4-p_5)^2+ \\
 &\phantom{\times \Big( (} (p_4-p_3)^2(p_2+p_5)^2(p_1-p_5)^2(p_3+p_5)^2\Big)
    \end{aligned}
\end{equation}
which, although rather huge, still has structure similar to the four-point case. Considering larger number of points will increase the number of terms, as they include cyclic permutations, but will not introduce anything conceptually different.
\section{Difference equations on string tree amplitudes}
In the previous sections, we have seen that the Veneziano amplitude satisfies a system of difference equations in the variables $\alpha s$ and $\alpha t$ which is a radical departure from the Feynman diagrams in the QFT, which typically have differential (or algebraic) relations. A natural generalization is to ask whether similar difference equations can be constructed for arbitrary tree amplitudes of an open bosonic string. \\
In this chapter, we will show that such equations do exist and can be obtained using two simple but powerful techniques. The first technique is the use of integration by parts (IBP) in integrals over the moduli of Riemann surfaces, which leads to relations between amplitudes with shifted arguments. The second technique is the so-called "trivial" identities that arise from replacing the unit by a sum of various combinations of coordinate differences $x_i-x_j$ this gives additional linear connections.\\
We will illustrate these methods by examples of 4-point and 5-point amplitudes, and then formulate a general scheme for an arbitrary number of external states. It turns out that the system of constructed difference equations contains exactly as many independent relations as there are kinematic parameters, which suggests the completeness of this system.
\begin{equation}
    A_n(p_1,\dots,p_n)=\int_{x_1<x_2<\dots<x_n}\frac{d^nx}{SL(2,R)}\frac{\prod_{i<j}(x_j-x_i)^{\alpha p_ip_j}}{(x_1-x_2)\dots(x_n-x_1)}
    \label{eq:general_tree_amp}
\end{equation}
\subsection{Four-point amplitude}
The simplest example is again the well-known four-point amplitude, namely:
\begin{equation}
    A_2(a,b)=\int_0^1dx x^{a-1}(1-x)^{b-1}
\end{equation}
where we introduced more convenient variables $a$ and $b$.\\
The common technique to get the identities on such kind of integrals is by putting the full derivative in the integrand and then proceeding with the integration by parts. Further on we will call the resulting expressions IBP identities. In this concrete case we will get the following:  
\begin{equation}
\begin{aligned}
    \int_{0}^1\frac{d(x^{a-1}(1-x)^{b-1})}{dx}dx &= (x^{a-1}(1-x)^{b-1})|^1_0 = 0\\
    \int_{0}^1\frac{d(x^{a-1}(1-x)^{b-1})}{dx}dx &= (a-1)A(a-1,b)-(b-1)A(a,b-1)
\end{aligned}
\end{equation}
It is convenient to shift the arguments, i.e. $a \rightarrow a + 1,\ b \rightarrow b + 1$, and write it with the use of difference operators to get the equation in more transparent form
\begin{equation}
\label{eq:four_point_ibp}
\boxed{\left(a D_b - b D_a + a - b\right)A_2(a,b)=0}
\end{equation}
However this gives us only one equation instead of the desired system of two. To get the remaining one we use different approach, i.e. the ``trivial'' one. Further we will somewhat distinguish this two ways of obtaining equations. The main essence of it is substituting $1=x+(1-x)$ or sum of any other combination of brackets in the integrand which gives 1. In this concrete case we use just $1=x+(1-x)$, thus obtaining the equation:
\begin{equation}
	\label{eq:four_point_trivial}
    \boxed{(D_a+D_b+1)A_2=0}
\end{equation}
Combining both the equations, i.e. \eqref{eq:four_point_trivial} and\eqref{eq:four_point_ibp}, we can get the system of equations on one variable each:
\begin{equation}
\begin{cases}
    \left(\left(a+b\right)D_a+b\right)A_2=0\\
    \left(\left(a+b\right)D_b+a\right)A_2=0
\end{cases}
\end{equation}
The resulting system is in the possible form, nevertheless this example, although trivial, demonstrates two main approaches to constructing difference equations on tree amplitudes we will further use.
\subsection{Five-point amplitude}
Let's proceed to the  more challenging case of 5 point amplitude, which can be expressed as the following integral:
\begin{equation}
    A_5(a,b,c,d,e)=\int_0^1dx\int_0^xdy(x^{a-1}y^{b-1}(1-x)^{c-1}(1-y)^{d-1}(x-y)^{e-1})
\end{equation}
Using substitutions $1 = x + (1-x)$, $1 = y + (1-y)$ and $1 = 1 + (x - y) - x + y$ one gets the system of 3 ``trivial'' equations:
\begin{equation}
\label{eq:five_point_trivial}
\boxed{
\begin{aligned}
    (D_a+D_c+1)A_5&=0\\
    (D_b+D_d+1)A_5&=0\\
    (D_e-D_a+D_b+1)A_5&=0
\end{aligned}
}
\end{equation}
Now we need to find only 2 equations, which we again will find using IBP. Putting the full derivative on $x$ and $y$ in the integrand we get the following relations on the amplitude:
\begin{equation}
\label{eq:five_point_ibp}
\begin{aligned}
    (a-1)A_5(a-1)-(c-1)A_5(c-1)+(e-1)A_5(e-1)&=0\\
    (b-1)A_5(b-1)-(d-1)A_5(d-1)-(e-1)A_5(e-1)&=0
\end{aligned}
\end{equation}
where we omitted all unchanged arguments of $A_5$, i.e. $A_5(a-1):=A_5(a-1,b,c,d,e)$ and etc.\\
However, in this case writing down these equation through the difference operators is rather challenging so we will write it down explicitly for the first relation in \eqref{eq:five_point_ibp} for more transparency.\\
First of all, let's do a shift of variables $a-1\to a,c-1\to c,e-1\to e$:
\begin{equation}
\label{eq:five_point_ibp_simplification}
    aA_5(c+1,e+1)-cA_5(a+1,e+1)+eA_5(a+1,c+1)=0
\end{equation}
Actually, we already can write it down in terms of difference operators now; however, we can further simplify it by expanding the terms in the following way:
\begin{equation}
\begin{aligned}
    A_5(c+1,e+1)&=/(1-x)(x-y)=(1-x)x-(1-x)y/=A_5(a+1,c+1)-A_5(c+1,b+1)\\
    A_5(a+1,e+1)&=/x(x-y)=x^2-xy/=A_5(a+2)-A_5(a+1,b+1)\\
    A_5(a+1,c+1)&=/x(1-x)=x-x^2/=A_5(a+1)-A_5(a+2)
\end{aligned}
\end{equation}
Plugging it in \eqref{eq:five_point_ibp_simplification} we get:
\begin{equation}
\label{eq:five_point_ibp_simplification_2}
    (a+e)A_5(a+1)-(a+e)A_5(a+2)-cA_5(a+2)-aA_5(c+1,b+1)+cA_5(a+1,b+1)=0
\end{equation}
The last thing to do is to express the last remaining term containing $c$ variable shift:
\begin{equation}
    A_5(c+1,b+1)=/(1-x)y=y-xy/=A_5(b+1)-A_5(a+1,b+1)
\end{equation}
And after the substitution in \eqref{eq:five_point_ibp_simplification_2} we arrive at:
\begin{equation}
    (a+e)A_5(a+1)-(a+e+c)A_5(a+2)-aA_5(b+1)+aA_5(a+1,b+1)+cA_5(a+1,b+1)=0
\end{equation}
Or in a more convenient difference operator form:
\begin{equation}
    \boxed{-(a+e+c)D^2_aA-(e+c)D_aA+cD_bA+(a+c)D_bD_aA=0}
\end{equation}
Similarly, we can process the second equation in \eqref{eq:five_point_ibp}:
\begin{equation}
    \boxed{(e+b+d)D^2_bA_5+(e+d)D_bA_5-dD_aA_5-(b+d)D_bD_aA_5=0}
\end{equation}
Thus, the resulting system reads:
\begin{equation}
\label{eq:5_point_system_init}
    \begin{cases}
        \left(D_a+D_c+1\right)A_5=0\\
        \left(D_b+D_d+1\right)A_5=0\\
        \left(D_e-D_a+D_b+1\right)A_5=0\\
        \left(-(a+e+c)D^2_a-(e+c)D_a+cD_b+(a+c)D_bD_a\right)A_5=0\\
        \left((e+b+d)D^2_b+(e+d)D_b-dD_a-(b+d)D_bD_a\right)A_5=0
    \end{cases}
\end{equation}
However, one can note that previously all the equations contained derivatives of a single variable each, providing simpler further analysis, while in this case they contain mixed derivatives. 
\subsection{General construction scheme}
As we have demonstrated earlier, it is possible to construct system of difference equations on tree string amplitudes using two simple ideas, namely:
\begin{eqnarray}
    1=(x_i-x_j) - (x_j-x_k) + (x_i-x_k) + 1 \rightarrow A_n= A_n(a_{i,j}+1) - A_n(a_{j,k}+1) + A_n(a_{i,k}+1) + A_n \label{eq:trivial_eqs}\\
    \int\limits_{x_1<x_2<\dots<x_n}\frac{d^nx}{SL(2,R)}\mathcolor{blue}{\frac{\partial}{\partial x_k}}\frac{\prod_{i<j}(x_j-x_i)^{\alpha p_ip_j}}{(x_1-x_2)\dots(x_n-x_1)}=0=\sum\limits_{j=1}^{n}(\alpha p_k p_j-1) A_n(a_{k,j}-1)\hspace{10mm} \label{eq:ibp_eqs}
\end{eqnarray}
where $a_{i,j}=\alpha p_i p_j$, i.e. scalar products of momenta on which the amplitude actually depends. Despite simplicity of those ideas they allow us to construct the system of equations which fixes the evolution of amplitude depending on $a, b, c$ and etc. It is not difficult to prove that this statement holds for n-point amplitude. n-point tree amplitude depends on $\frac{n(n-3)}{2}$ number of parameters, while the proposed construction scheme produces
\begin{equation}
    \begin{aligned}
        &\eqref{eq:trivial_eqs} \rightarrow \frac{(n-2)(n-3)}{2}\ \text{equations}\\
        &\eqref{eq:ibp_eqs} \rightarrow \hspace{8mm} n-3 \hspace{9mm} \text{equations}
    \end{aligned}
\end{equation}
The acquired system of equations consists of independent equations and their number is exactly the same as the number of variables in n-point amplitude. However it still remains an open question if those equations define \textit{only} the string amplitude in a sense that there can be other solutions to this system. We leave this question out of scope of this paper. 

\section{The $\alpha\longrightarrow 0$ limit}
In this section, we analyze the low-energy regime of our framework, focusing on the limit in which the $\alpha\to0$. This limit plays a central role in clarifying the structure and physical interpretation of the equations, as it isolates the dominant contributions relevant at low energies. To further illustrate the practical implementation and consequences of this limiting procedure, we examine a series of representative toy models. These examples are chosen to highlight the behavior of the system under controlled simplifications, providing both intuition and concrete demonstrations of how the limit operates within our formalism.
\subsection{Toy Example}
We begin by considering a simplified toy model of the Veneziano amplitude that retains the essential structural features of the full expression. In particular, this model is constructed to satisfy a class of finite-difference equations analogous to those obeyed by the exact amplitude. Studying this reduced setting allows us to isolate the key mechanisms underlying the difference relations while keeping the analysis technically transparent.
\begin{equation}
    A^{Toy}(s,t)=\int_0^1dxx^{\alpha s-1}(1-x)^{\alpha t-1}=\frac{\Gamma(\alpha s)\Gamma(\alpha t)}{\Gamma(\alpha s+\alpha t)}\to(\alpha\to0)\to\frac{1}{s}+\frac{1}{t}
\end{equation}
This amplitude satisfies a system of difference equations based on the shifts of the kinematic variables:
\begin{equation}
\label{eq:beta}
    \begin{aligned}
      &((s+t)D_{s}+t)A^{Toy}(s,t)=0 \\
      & ((s+t)D_t+s)A^{Toy}(s,t)=0
    \end{aligned}
\end{equation}
Consider the limit $\alpha\to 0$, in this regime, the shift amplitude $1/\alpha$ becomes parametrically large, while the reference kinematic point $(s,t)$ remains fixed. Under these conditions, the shifted contribution directly describes the asymptotic sector of the amplitude. It is important that the selection of this sector does not require additional scaling of the local kinematic variables, since the asymptotic behavior arises solely due to the growth of the shear parameter.\\
We assume that the amplitude admits an asymptotic stabilization with respect to large shifts in the s-channel:
\begin{equation}
    A^{Toy}(s+\frac{1}{\alpha},t)=f(t)+o(1),\alpha\to0
\end{equation}
uniformly for fixed s in the kinematic domain of interest. Physically, this expresses that sufficiently large Regge–type displacements drive the system toward an effective particle configuration governed only by the complementary channel variable.\\
Using our assumption on our equations. we can get an asymptotic expression:
\begin{equation}
    \begin{aligned}
        &sA^{Toy}(s,t)=f(t)(s+t) \\
        &tA^{Toy}(s,t)=g(s)(s+t)
    \end{aligned}
\end{equation}
By solving the system, we obtain the relation between the functions $g(s)=1/s,f(t)=1/t$, resulting in an algebraic equation that is consistent with our QFT:
\begin{equation}
    stA^{Toy}(s,t)=s+t
\end{equation}
\subsection{Two Beta functions}
Before moving on to more complex structures, it is useful to consider a simple model consisting of the sum of two beta functions:
\begin{equation}
    A^{Toy}_2(s,t)=B(s,t)+B(s+1,t)
\end{equation}
The reason for considering this particular combination is fundamental. In string theory, all four-point amplitudes at the tree level, regardless of the type of theory (bosonic, superstring, or photonic), are expressed as linear combinations of shifted beta functions. Thus, having a way to construct equations on them is of the particular interest.\\
First of all we need to write down the system of equations, which nullify this function and there are two possible ways of doing so. We can use the properties of beta functions and derive them straightforwardly or use the already found out one for single beta function \eqref{eq:beta}. Here we will choose the second option as it is more general and depend only on the initial equation, i.e equation on one beta function in this particular case, and some ``shift'' operator \eqref{eq:two_beta_shift}.\\
To do it we need to express $A^{Toy}_2$ as the action of some difference operator on initial beta function:
\begin{equation}
\label{eq:two_beta_shift}
    A^{Toy}_2 = (D_s+2)B(s,t)
\end{equation}
This allows to express $D_s$ and $D_t$ of $A^{Toy}_2$ in terms of derivatives of $B(s,t)$, for which we already have the equations they satisfy.\\
The only thing left is to compute $D_s A^{Toy}_2, D_s^2 A^{Toy}_2, \dots$ and for $t$-channel respectively till these derivatives become linear dependent. In this naive example we will need only $D_s A^{Toy}_2$ and $D_t A^{Toy}_2$. As the initial system is of the first order, we can express all the derivatives in terms of $B(s,t)$:
\begin{equation}
    \begin{aligned}
    	&A^{Toy}_2 = (D_s+2)B(s,t) = /\eqref{eq:beta}/ = \frac{2s+t}{s+t}B(s,t)\\
        &D_s A^{Toy}_2 = (D_s^2 + 2 D_s)B(s,t) = /\eqref{eq:beta}/ = -\frac{t(2s+t+1)}{(s+t)(s+t+1)}B(s,t)\\
        &D_t A^{Toy}_2 = (D_t D_s + 2 D_t)B(s,t) = /\eqref{eq:beta}/= -\frac{s(2s+t+2)}{(s+t)(s+t+1)}B(s,t)
    \end{aligned}
\end{equation}
Thus we can construct nullifying linear combinations of these functions for $s-$ and $t-$ channels respectively:
\begin{equation}
\label{eq:two_beta}
\begin{aligned}
	&\Big(\left(2s+t\right)\left(s+t+1\right)D_s+t\left(2s+t+1\right)\Big)A^{Toy}_2=0\\
	&\Big(\left(s+t+1\right)\left(2s+t\right)D_t+s\left(2s+t+2\right)\Big)A^{Toy}_2=0
\end{aligned}
\end{equation}
Again, as in the previous section for the case of one beta function, we put
\begin{equation}
A^{Toy}_2 (s+\frac{1}{\alpha},t)=f(t)+o(1),\ \alpha\to0
\end{equation}
and for the $t-$ channel respectively. Thus \eqref{eq:two_beta} transforms to
\begin{equation}
\label{eq:two_beta_subfinal}
\begin{aligned}
A^{Toy}_2 (s,t) & = \frac{(s+t+1)(2s+t)}{t(2s+t+1)} g(s) \\
A^{Toy}_2 (s,t) & = \frac{(2s+t)(s+t+1)}{s(2s+t+2)} f(t)
\end{aligned}
\end{equation}
Making an expansion on $\alpha\to 0$ we can find out the value of $g(s)$ and $f(t)$ in leading order up to the common constant multiplier:
\begin{equation}
g(s) = \frac{C}{s} + O(1), \hspace{5mm} f(t) = \frac{2C}{t} + O(1), \hspace{5mm} C\in\mathbb{C}
\end{equation}
Putting it back at \eqref{eq:two_beta_subfinal} and imposing $C=1$ we get
\begin{equation}
A^{Toy}_2 (s,t) = \frac{1}{s} + \frac{2}{t} + O(1),\ \alpha\to 0
\end{equation}
as it should be.\\
This algorithm allows straightforward generalization to more analytically challenging amplitudes, where working with functional equation, unlike in the beta function case, becomes difficult, as it depends only the differential equation and nothing else. Nevertheless, it still allows to reproduce the exact answers, as finding solutions for such equations is straightforward.
\section{Further research}
$\textbf{Mellin-transform}$: The difference equations for tree string amplitudes obtained in the previous sections are naturally reformulated in terms of Mellin transforms. This transformation not only simplifies the structure of the equations (transforming difference operators into shift operators), but also establishes a connection with the theory of zeta functions and their generalizations. In a recent paper \cite{remmen2026zetafunctionssuperstring} the Mellin transform of the four-point string amplitude was proposed:
\begin{equation}
    \Omega(z,t)=\frac{1}{2\pi i}\oint_{C}\frac{ds}{s^z}A(s,t)
\end{equation}
where the contour C encloses the poles of the amplitude A(s,t) in the complex plane s. When t=0 this expression reduces to the Riemann zeta function $\zeta(z)$ and when $t\neq0$ it gives its non-trivial deformation. In the same work, some shift relations for $\Omega(z,t)$ are also present.\\
Our approach allows us to obtain these relations systematically, starting from difference equations, and, more importantly, to generalize them to the case of an arbitrary number of external particles.
\begin{eqnarray}
    ((s+t)D_t+s)A^{Toy}(s,t)=0 \xrightarrow[]{\text{Mellin transform}} tD_t\Omega(z,t)+\Omega(z-1,t+1)=0
\end{eqnarray}
Our algorithm for constructing difference equations allows us to study the shift relations on the multidimensional Mellin image
\begin{equation}
 \Omega_n(z_1,...,z_{n-1},t)=\frac{1}{(2\pi i)^{n-3}}\oint_{C_1}...\oint_{C_{n-1}}\frac{ds_1...ds_{n-1}}{s_1^{z_1}...s_{n-1}^{z_{n-1}}}A_n(s_1,...,s_{n-1},t)
\end{equation}
which paves the way for studying the multidimensional deformations of zeta functions.

\textbf{Other string theories}: The presented algorithm for constructing difference equations relies solely on the structure of integrals over the moduli space and does not use the specifics of the bosonic string in 26 dimensions. This allows us to hope for its direct generalization to other classes of string theories. The most direct candidates are superstrings and closed strings. As is known, the amplitudes of closed strings at the tree level can be obtained from the amplitudes of open strings by means of the KLT relations \cite{Kawai:1985xq}, which suggests the possibility of transferring the difference equations to this case as well.

This would allow us to determine whether the property of satisfying difference equations is a universal feature of all string amplitudes, or whether it is specific to the bosonic string.

\textbf{Hierarchy of equations}: One of the naturally arising questions is whether difference equations are a universal characteristic of all tree-level string amplitudes. This would allow us to consider the entire string theory at the tree level as a unified hierarchical structure, where the fundamental objects, the amplitudes, are defined not by integrals, but by systems of difference equations. This approach potentially can shed new light on string theory and lead to a deeper understanding of its internal structure.

\textbf{One-loop string amplitude}: This work lays the foundations for a systematic description of tree-level string amplitudes using difference equations. A natural and important extension is to generalize this formalism to the loop level. In quantum field theory loop diagrams satisfy the Picard–Fuchs differential equations, whose coefficients are rational functions of kinematic invariants. This is a direct consequence of the fact that Feynman integrals are believed to be periods of algebraic varieties. In string theory, a similar structure is expected to be much richer.

Recent results \cite{Eberhardt_2023,Eberhardt_2023_2} provide an explicit integral representation for the imaginary part of type I and type II one-loop amplitudes. The structure suggests that after summation and integration, the amplitude will satisfy a system of linear differential equations in s and t. The coefficients of these equations, however, will not be rational, which reflects a non-trivial dependence on the torus modulus. In the limit $\alpha\to 0$ these coefficients should turn into rational functions, and the equations themselves should turn into the well-known Picard–Fuchs equations for the corresponding Feynman diagrams \cite{Mishnyakov:2024xjz,Lairez:2022zkj,Mishnyakov:2024rmb}.

\section{Conclusion}
In this work, we suggest to begin the study of string amplitudes as $D$-modules -- solution of the system of
Picard-Fucks-like equations for integrals over moduli spaces -- 
and develop this into a general formalism behind the similar approach to Feynman integrals in QFT,
which attracts a lot of attention in recent years.
We begin from consideration of tree string amplitudes, where the derivation of
Picard-Fuchs equations is already non-trivial -- what is not so obvious at CFT level,
where integration over moduli space can be completely eliminated, 
say, by working in coordinate space like in \cite{Mishnyakov:2024rmb}.
Equations for string amplitudes are, however, less simple and we consider them and their particle-theory limits
in some detail.
Further extension to loop diagrams would give a  hope for a complete classification of string amplitudes,
based on the theory of functional equations, 
which is an ambitious and promising direction in modern string theory.

\section*{Acknowledgements}

Funding for this publication was generously provided by the Priority 2030 Academic Leadership Initiative, contributing
to the educational work of "Universities for a New Generation of Leaders", a project within the framework of the federal Youth and Children program.

\printbibliography

\end{document}